\documentclass[onecolumn]{aastex61}
\def\robbie{{\sc Robbie}}
\def\fitswarp{{\sc Fits\_Warp}}
\def\aegean{{\sc Aegean}}

\begin{document}

\title{Multi-epoch Low Radio Frequency Surveys of the {\it Kepler K2} Mission Campaign Fields 3, 4, and 5 with the Murchison Widefield Array}

\author[0000-0002-8195-7562]{Tingay, S.J.}
\affiliation{International Centre for Radio Astronomy Research, Curtin University, Bentley, WA 6102, Australia}

\author[0000-0002-4203-2946]{Hancock, P.J.}
\affiliation{International Centre for Radio Astronomy Research, Curtin University, Bentley, WA 6102, Australia}

\begin{abstract}
We present Murchison Widefield Array (MWA) monitoring of the Kepler K2 mission Fields 3, 4, and 5 at frequencies of 155 and 186 MHz, from observations contemporaneous with the K2 observations.  This work follows from previous MWA and GMRT surveys of Field 1, with the current work benefiting from a range of improvements in the data processing and analysis.  We continue to build a body of systematic low frequency blind surveys overlapping with transient/variable survey fields at other wavelengths, providing multi-wavelength data for object classes such as flare stars.  From the current work, we detect no variable objects at a surface density above $2\times10^{-4}$ per square degree, at flux densities of $\sim 500$\,mJy, and observation cadence of days to weeks, representing almost an order of magnitude decrease in measured upper limits compared to previous results in this part of observational parameter space.  This continues to show that radio transients at metre and centimetre wavelengths are rare.
\end{abstract}


\section{Introduction}
The Kepler K2 mission commenced in 2014, designed to use the Kepler spacecraft with a reduced number of reaction wheels, to observe a number of fields near the ecliptic \citep{2014PASP..126..398H}.

Previously, we undertook a pilot study of the K2 Field 1 using the Murchison Widefield Array (MWA: \citet{2013PASA...30....7T}) and the TGSS ADR1 \citep{2017A&A...598A..78I} using the Giant Metre-wave Radio Telescope (GMRT: \citet{1991ASPC...19..376S}) to monitor the field at low frequencies (140 - 200 MHz) during the K2 campaign \citep{PaperI}.  We produced a source catalog for this field containing over 1000 radio sources.  No transient or variable radio sources were found during the Field 1 MWA campaign; the catalog produced a unique supporting resource for K2 mission science.

Using the lessons learned during the Field 1 MWA observations and subsequent data processing, we have now undertaken further monitoring observations of K2 Fields 3, 4, and 5 concurrent with the K2 observations.  We report these observations in this paper.  We used an improved scheduling strategy for these new observations, allowing a more robust and accurate extraction of radio light curves from our data.  Thus, our sensitivity to transients and variables is improved, relative to the results reported in \citet{PaperI}.

The K2 mission is making significant strides in the direction blazed by Kepler, in the detection and study of large numbers of exoplanets.  However, the science goals of K2 observations are widely varied.  Of relevance to the results of our MWA monitoring of K2 fields, a range of work aimed at studying the optical variability of active galactic nuclei (AGN) is starting to emerge \citep{2017AAS...22920302S,2018AAS...23125021W,2017AAS...22925039C}.  K2 has also detected M dwarf flare stars that require study at radio frequencies, in particular low frequencies where such stars are most active \citep{2018AAS...23133407V}.

A search for M dwarf flares was performed with our previously published data for K2 Field 1, adding to a growing literature in this field that is relevant to exoplanet habitability research.  In particular, our previous K2 Field 1 results and the MWA results of \citet{2016MNRAS.458.3506R} did not detect any M dwarf flares.  However, using the MWA to measure circular polaristion and therefore achieve higher sensitivity to flares, \citet{2017ApJ...836L..30L} found intermittent flares from targeted observations of UV Ceti, but not from YZ CMi or CN Leo.  Only sparse observational data exist at MWA frequencies for M dwarf flare stars and further systematic monitoring studies are required, as well summarised in \S5.3.3 of the review by \citet{2019PASP..131a6001M}.  The TESS mission \citep{2015JATIS...1a4003R} is being used to monitor large numbers of M dwarf flares \citep{2019arXiv190100443G} during the Cycle 1 portion of the mission, covering southern declinations well matched to follow-up with MWA observations.

While various classes of radio transients are expected to be rare at centimeter and meter wavelengths \citep{2015ApJ...806..224M}, significant work is being expended testing these limits \citep{2014MNRAS.438..352B,2016MNRAS.459.3161C,2013ApJ...762...93C,2015MNRAS.446.2560M,2015JAI.....450004O,2016ApJ...832...60P,2016MNRAS.458.3506R,2017MNRAS.466.1944M} with existing large-scale instruments and detecting occasional interesting transients and variables \citep{2005Natur.434...50H,2009ApJ...696..280H,2012AJ....143...96J,2017MNRAS.466.1944M}.  Much of this work is in preparation for transient research with the Square Kilometre Array \citep[SKA,][]{2015aska.confE..51F}.

Often through radio transient surveys, once detections are made, the follow-up identification, verification, and interpretation of the objects is difficult.  The goal of the observations reported here is to continue to build unbiased, wide-field, multi-epoch surveys at low radio frequencies, covering the full extent of the K2 fields concurrent with the optical monitoring observations, to facilitate the multi-wavelength search for transients and variables.  In \S2 we present the observations and data processing, including improvements over \citet{PaperI}.  In \S3 we present our results, and we discuss them and our conclusions in \S4.

\section{Observations and Data Processing}

\subsection{Observations}
The parameters of the MWA observations conducted for K2 fields 3, 4, and 5 are given in Table \ref{tab1}, for the period 2014 November 07 to 2015 May 28.  All observations were performed in a standard imaging mode, as described in \citet{PaperI}, at the same center frequencies of 154.88 MHz and 185.60 MHz and with the same processed bandwidth of 30.72 MHz (24$\times$1.28 MHz coarse channels, each comprised of 128$\times$10 kHz fine channels).

However, the scheduling of these observations benefited from our prior experiences recorded in \citet{PaperI}, in that we used a single az/el pointing for each of the three fields of interest. This guarantees that all observations for a given field have the same LST, and thus a primary beam and synthesized beam that are consistent throughout the observation sequence.  This results in data that are simpler to process in terms of extracting light curves and more robust light curves due to smaller errors on individual flux densities.

We recovered the MWA data from the new MWA All-Sky Virtual Observatory (ASVO) node\footnote{https://asvo.mwatelescope.org/}, which allows users to discover observations and download the visibility and calibration data in a variety of formats, applying user-defined manipulations to the visibilities.  We downloaded the observations described in Table \ref{tab1} using an averaging time of 2 seconds, a frequency averaging of 40 kHz (corresponding to four fine channels), and flagging 160 kHz at the edges of coarse channels (corresponding to 16 fine channels at each coarse channel edge).  Recorded flagging information was also applied, as was default radio frequency interference (RFI) flagging via AO Flagger \citep{2012MNRAS.422..563O}.  The visibility data were extracted as uvfits files, suitable for import into MIRIAD \citep{1995ASPC...77..433S}.

The total volume of MWA visibility data processed (including calibration data) was approximately 4.4 TB, with the visibility averaging parameters described above applied during data extraction from the ASVO.

For Field 3, 36 observations were scheduled, but only 18 were successful.  18 observations between 2014 November 25 and 2014 December 13 were affected by a lightning strike that caused a power outage at the MWA (completely lost four observations) and caused temporary damage to 43/128 tiles (substantially degraded data).  Thus, only the 18 successful observations are listed in Table \ref{tab1} and described in the remainder of the paper.

Other issues with data from other Fields are described in the following section, primarily regarding challenges introduced by the far northern pointings and strong sources in the high response primary beam sidelobes.  Any future monitoring of K2 fields with the MWA could concentrate on fields south of the equator, to minimise such issues.

\startlongtable
\begin{deluxetable}{c c c c c r r}
\tablecaption{\label{tab1}MWA observation log}
\startdata
OBSID & START DATE/TIME& T& TARGET & FREQ& RA& DEC \\ 
&(UT)&(s)&&(MHz)&($^{\circ}$)&($^{\circ}$) \\ \hline
1099403792 & 2014-11-07 13:56:16 & 296 & F3 & 154.88 & 334.454 & -14.675 \\
1099392080 & 2014-11-07 10:41:03 & 176 & 3C444 & 154.88 & 330.524 & -19.773 \\
1099404096 & 2014-11-07 14:01:19 & 296 & F3 & 185.60 & 335.724 & -14.676 \\
1099392256 & 2014-11-07 10:43:59 & 176 & 3C444 & 185.60 & 331.26 & -19.774 \\
1099576120 & 2014-11-09 13:48:24 & 296 & F3 & 154.88 & 334.453 & -14.675 \\
1099478520 & 2014-11-08 10:41:43 & 176 & 3C444 & 154.88 & 331.678 & -19.774\\
1099576424 & 2014-11-09 13:53:27 & 296 & F3 & 185.60 & 335.723 & -14.676 \\
1099478704 & 2014-11-08 10:44:48 & 176 & 3C444 & 185.60 & 332.447 & -19.775 \\
1099748448 & 2014-11-11 13:40:32 & 296 & F3 & 154.88 & 334.452 & -14.675 \\
1099910736 & 2014-11-13 10:45:19 & 176 & 3C444 & 154.88 & 330.274 & -19.968 \\
1099748752 & 2014-11-11 13:45:35 & 296 & F3 & 185.60 & 335.723 & -14.676 \\
1099910920 & 2014-11-13 10:48:24 & 176 & 3C444 & 185.60 & 331.043 & -19.968 \\
1099920776 & 2014-11-13 13:32:40 & 296 & F3 & 154.88 & 334.452 & -14.675 \\
1099910736 & 2014-11-13 10:45:19 & 176 & 3C444 & 154.88 & 330.274 & -19.968 \\
1099921080 & 2014-11-13 13:37:43 & 296 & F3 & 185.60 & 335.722 & -14.676 \\
1099910920 & 2014-11-13 10:48:24 & 176 & 3C444 & 185.60 & 331.043 & -19.968 \\
1100093104 & 2014-11-15 13:24:48 & 296 & F3 & 154.88 & 334.451 & -14.675 \\
1100083632 & 2014-11-15 10:46:55 & 176 & 3C444 & 154.88 & 332.647 & -19.969 \\
1100093408 & 2014-11-15 13:29:51 & 296 & F3 & 185.60 & 335.722 & -14.676 \\
1100083808 & 2014-11-15 10:49:51 & 176 & 3C444 & 185.60 & 333.383 & -19.969 \\
1100265432 & 2014-11-17 13:16:56 & 296 & F3 & 154.88 & 334.45 & -14.675 \\
1100256520 & 2014-11-17 10:48:24 & 176 & 3C444 & 154.88 & 334.987 & -19.97 \\
1100265736 & 2014-11-17 13:21:59 & 296 & F3 & 185.60 & 335.721 & -14.676 \\
1100256704 & 2014-11-17 10:51:27 & 176 & 3C444 & 185.60 & 335.756 & -19.971 \\
1100437760 & 2014-11-19 13:09:04 & 296 & F3 & 154.88 & 334.45 & -14.675 \\
1100429416 & 2014-11-19 10:50:00 & 176 & 3C444 & 154.88 & 330.124 & -19.773 \\
1100438064 & 2014-11-19 13:14:07 & 296 & F3 & 185.60 & 335.72 & -14.676 \\
1100429592 & 2014-11-19 10:52:56 & 176 & 3C444 & 185.60 & 330.859 & -19.773 \\
1100610096 & 2014-11-21 13:01:20 & 296 & F3 & 154.88 & 334.483 & -14.675 \\
1100602304 & 2014-11-21 10:51:27 & 176 & 3C444 & 154.88 & 332.464 & -19.774 \\
1100610392 & 2014-11-21 13:06:16 & 296 & F3 & 185.60 & 335.72 & -14.676 \\
1100602488 & 2014-11-21 10:54:31 & 176 & 3C444 & 185.60 & 333.233 & -19.775 \\
1100782416 & 2014-11-23 12:53:19 & 296 & F3 & 154.88 & 334.449 & -14.675 \\
1100775200 & 2014-11-23 10:53:04 & 176 & 3C444 & 154.88 & 334.837 & -19.776 \\
1100782720 & 2014-11-23 12:58:24 & 296 & F3 & 185.60 & 335.719 & -14.676 \\
1100775376 & 2014-11-23 10:56:00 & 176 & 3C444 & 185.60 & 335.573 & -19.776 \\ \hline
1108551264 & 2015-02-21 10:54:08 & 296 & F4 & 154.88 & 56.437 & 19.719 \\
1108734680 & 2015-02-23 13:51:04 & 112 & HydA & 154.88 & 138.646 & -11.143 \\
1108551560 & 2015-02-21 10:59:03 & 296 & F4 & 185.60 & 57.677 & 19.721 \\
1108734808 & 2015-02-23 13:53:11 & 112 & HydA & 185.60 & 139.181 & -11.143 \\
1108725688 & 2015-02-23 11:21:12 & 296 & F4 & 185.60 & 57.344 & 21.244 \\
1108734808 & 2015-02-23 13:53:11 & 112 & HydA & 185.60 & 139.181 & -11.143 \\
1109070040 & 2015-02-27 11:00:23 & 296 & F4 & 154.88 & 56.073 & 21.243 \\
1108986336 & 2015-02-26 11:45:20 & 112 & HydA & 154.88 & 135.448 & -11.788	\\
%
%
\hline
1114081920 & 2015-04-26 11:11:43 & 296 & F5 & 154.88 & 131.325 & 18.973 \\
1114036848 & 2015-04-25 22:40:32 & 176 & 3C444 & 154.88 & 332.197 & -18.177 \\
1114082216 & 2015-04-26 11:16:39 & 296 & F5 & 185.60 & 132.563 & 18.974 \\
1114037024 & 2015-04-25 22:43:28 & 176 & 3C444 & 185.60 & 332.933 & -18.177 \\
1114254248 & 2015-04-28 11:03:52 & 296 & F5 & 154.88 & 131.325 & 18.973 \\
1114209712 & 2015-04-27 22:41:36 & 176 & 3C444 & 154.88 & 334.437 & -18.178 \\
1114254544 & 2015-04-28 11:08:47 & 296 & F5 & 185.60 & 132.562 & 18.974 \\
1114209888 & 2015-04-27 22:44:32 & 176 & 3C444 & 185.60 & 335.172 & -18.179 \\
1114598904 & 2015-05-02 10:48:08 & 296 & F5 & 154.88 & 131.324 & 18.973 \\
1114555432 & 2015-05-01 22:43:35 & 176 & 3C444 & 154.88 & 331.437 & -19.186 \\
1114599200 & 2015-05-02 10:53:04 & 296 & F5 & 185.60 & 132.561 & 18.974 \\
1114555616 & 2015-05-01 22:46:39 & 176 & 3C444 & 185.60 & 332.206 & -19.187 \\
1114943560 & 2015-05-06 10:32:24 & 296 & F5 & 154.88 & 131.322 & 18.973 \\
1114901160 & 2015-05-05 22:45:44 & 176 & 3C444 & 154.88 & 335.916 & -19.189 \\
1114943856 & 2015-05-06 10:37:20 & 296 & F5 & 185.60 & 132.559 & 18.974 \\
1114901344 & 2015-05-05 22:48:48 & 176 & 3C444 & 1855.60 & 336.686 & -19.19 \\
1115632872 & 2015-05-14 10:00:56 & 296 & F5 & 154.88 & 131.32 & 18.973 \\
1115592624 & 2015-05-13 22:50:07 & 176 & 3C444 & 154.88 & 330.366 & -19.971 \\
1115805200 & 2015-05-16 09:53:04 & 296 & F5 & 154.88 & 131.319	& 18.973 \\
1115765488 & 2015-05-15 22:51:11 & 176 & 3C444 & 154.88 & 332.606 & -19.972 \\
1115805504 & 2015-05-16 09:58:07 & 296 & F5 & 185.60 & 132.59 & 18.974 \\
1115765664 & 2015-05-15 22:54:07 & 176 & 3C444 & 185.60 & 333.342 & -19.973 \\
\enddata
\tablecomments{Column 1 - MWA observation ID; Column 2 - UT date and time of observation start; Column 3 - duration of observation in seconds; Column 4 - observation target (F3 = Field 3; F4 = Field 4; F5 = Field 5; HydA = Calibrator Hydra A; 3C 444 = Calibrator 3C444); Column 5 - Centre frequency in MHz; Column 6 - Right Ascension in decimal degrees; and Column 7 - Declination in decimal degrees.  The full observation table is available as a Machine Readable Table (CSV format).}
\end{deluxetable}

\subsection{Data Processing}

\subsubsection{Calibration and Imaging}\label{sec:imaging}

Data processing proceeded as described in Sections 3.1.1 and 3.1.2 in \citet{PaperI} and readers are referred to those descriptions for details.  An additional calibrator was used for our processing for K2 fields 3, 4, and 5, 3C 444.  In cases where 3C 444 is the calibrator, the model for the flux density was parameterized as $79.6\frac{\nu}{160 MHz}^{-0.88}$ Jy and a restriction in self-calibration was applied when imaging the calibrator, such that only baselines in the uv range 0.05 to 0.8 k$\lambda$ were used against this model.

\begin{figure}[h]
\begin{centering}
\includegraphics[height=0.3\textheight]{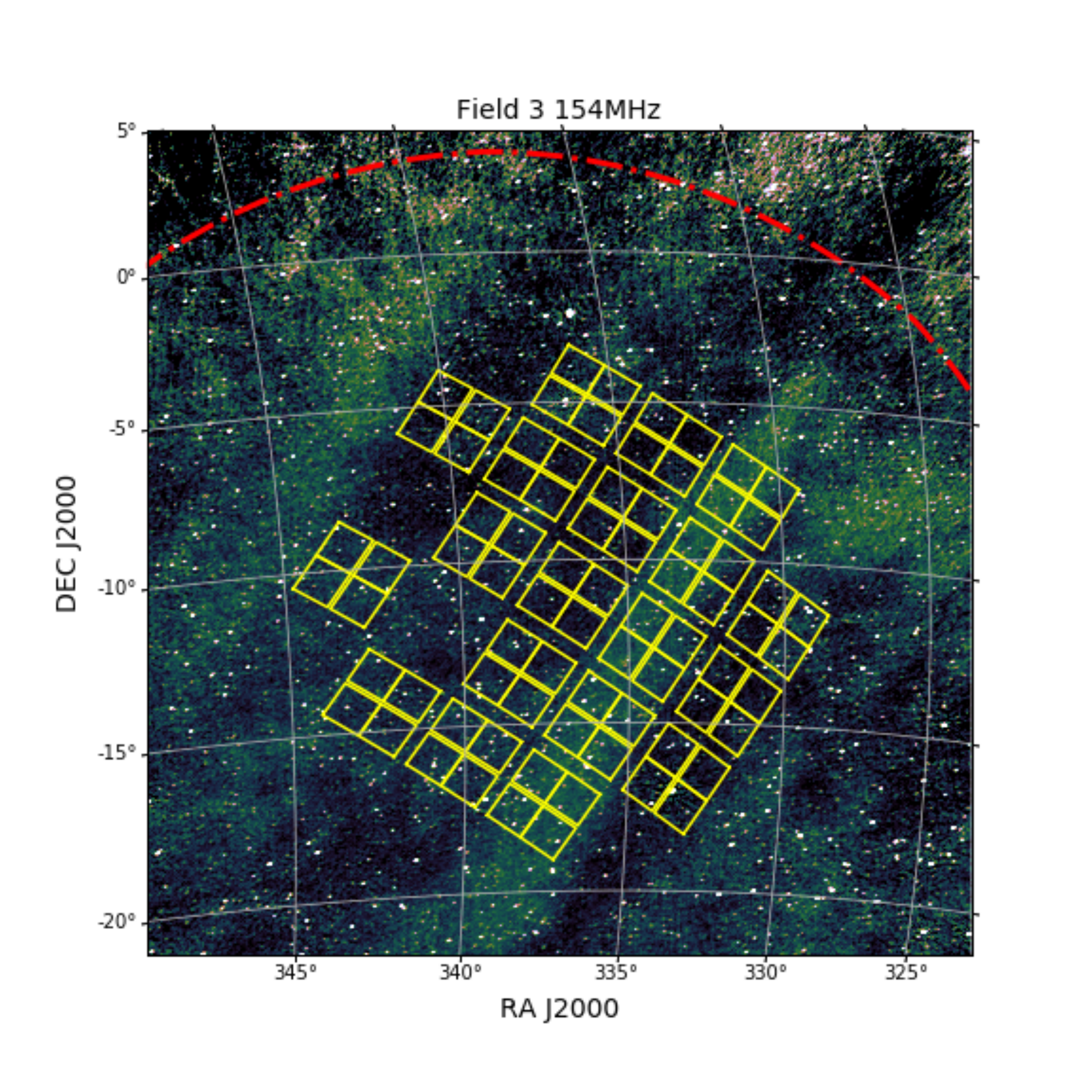}
\includegraphics[height=0.3\textheight]{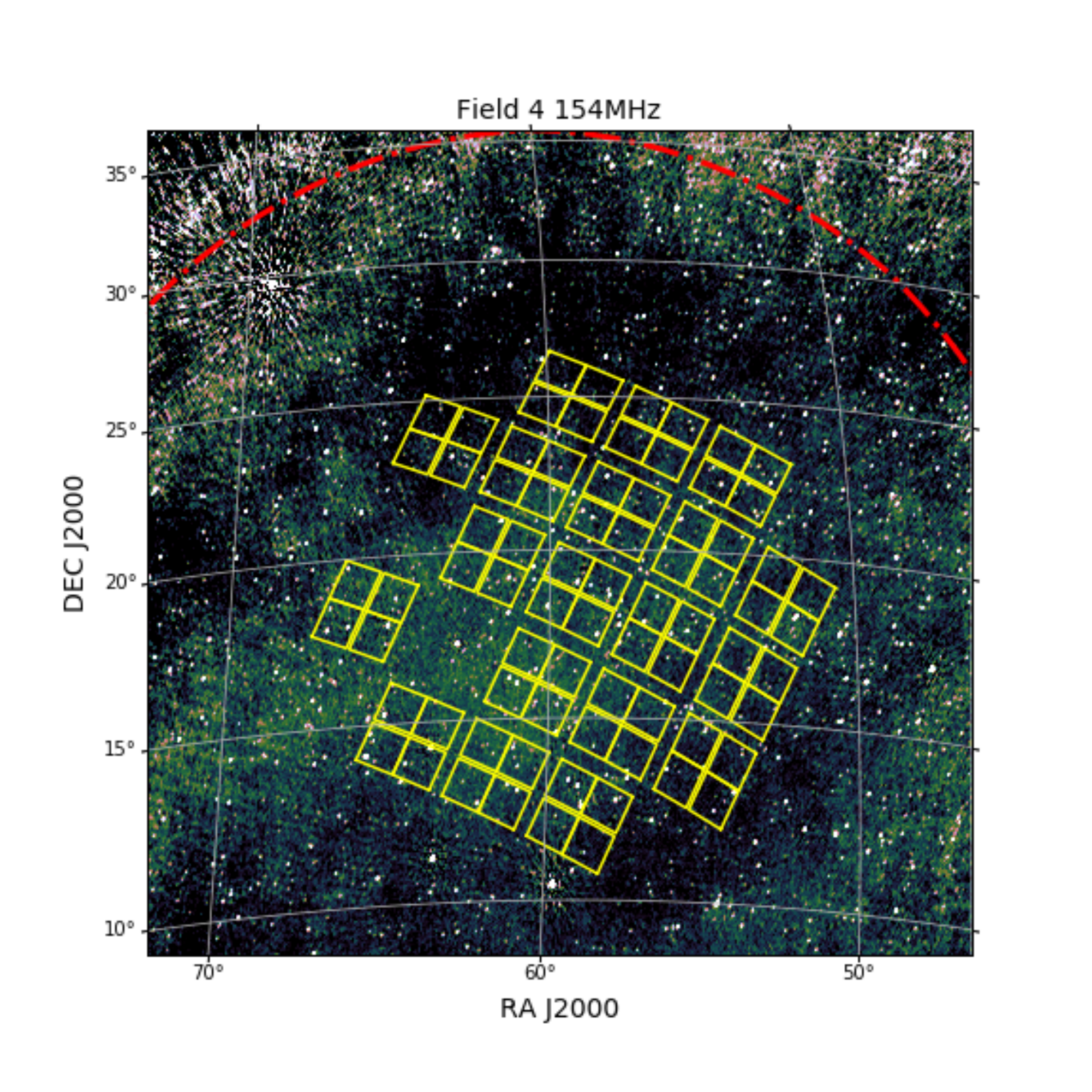}
\includegraphics[height=0.3\textheight]{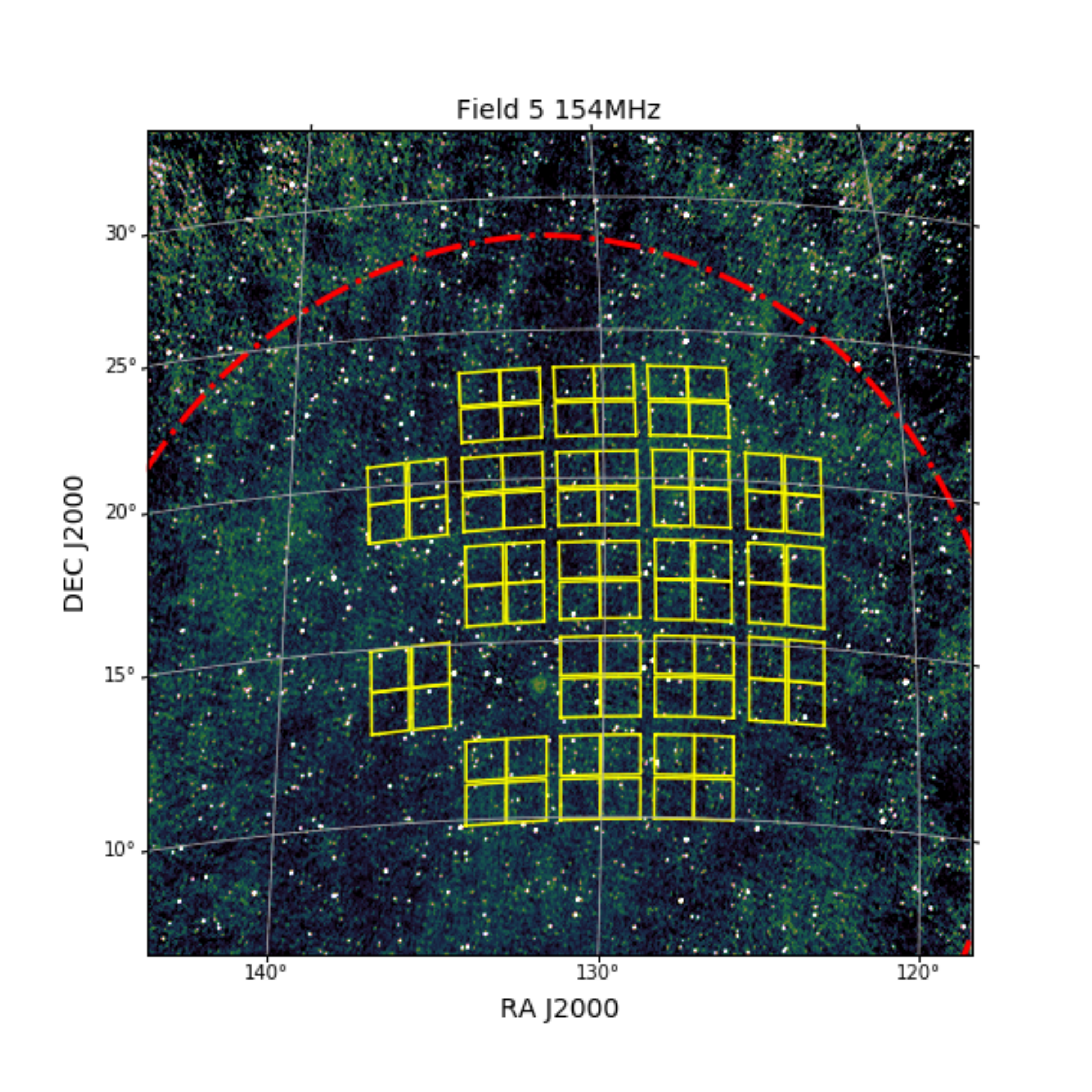}
\caption{
The three K2 fields monitored for variable and transient radio sources.
Images shown are typical of those produced at the two frequencies, using the procedure described in the text, and have been cropped to show just the region covered by the Kepler sensors.
The intensity scales range from $-0.2$ to $1$\,Jy/beam. 
The K2 areal footprint is shown in a yellow overlay.
The dot-dashed red circle shows the exclusion zone chosen by \robbie{}.}
\label{Fig1} 
\label{fig:fields}
\end{centering}
\end{figure}

Fields 4 and 5 are at significant positive declinations, $\sim+19^{\circ}$ and $\sim+17^{\circ}$, respectively, compared to Field 3 at $\sim-11^{\circ}$.  Imaging and calibration proved less consistently of high quality for Fields 4 and 5 than for Field 3.  While all three fields provided high quality images at some epochs (examples shown in Figure \ref{Fig1}), it was noted that a significant fraction of epochs for Fields 4 and 5 produced unusable images with the procedure described in \citet{PaperI}.  In some of these cases, we were able to recover images of a quality to be usable by removing from our imaging and calibration pipeline the final self-calibration step that included amplitude self-calibration.  

Possible reasons for Fields 4 and 5 being more challenging for imaging and calibration could be ionospheric activity through a greater path length at low elevations, radio frequency interference at lower elevation observations, or the presence of nearby bright radio sources in the primary beam sidelobes (e.g. the Crab Nebula, the Southern Galactic Plane, and the Sun, in various combinations), which have a high response for northern pointing directions.  An inspection of the logs recording the instrument status at these epochs reveals no significant issues with the intrinsic data quality. 

In some cases, the imaging and calibration process did not converge, after attempting a range of variations on the pipeline.  In particular, no epoch for Field 4 in the month of 2015 March was recoverable.  During those epochs, we noted that the far northern pointing produced sidelobes with a peak response of $>$0.5 of the main beam, meaning comparable power was contained in the visibilities from the opposite side of the sky.  In addition, the Sun was above the horizon at increasing elevations during the month.  This combination makes reliable imaging and calibration extremely difficult and the data were discarded.
The set of usable data are listed in Table\,\ref{tab1}.

\subsubsection{Source Finding and Production of Light Curves}\label{sec:sourcefinding}

The images described in \S\,\ref{sec:imaging} were inspected manually to remove any which were not usable.
The (effectively full-sky) images were then cropped to contain only a region of interest which was determined to be the largest region over which we could obtain accurate flux measurements.
The usable area of Field 5 is less than that of Fields 3 and 4 due to the presence of the Galactic plane in one of the primary beam side-lobes which increased the image noise away from the beam center.
The images were then processed using \robbie{}\footnote{www.github.com/PaulHancock/Robbie} - a work-flow for detecting and characterizing variable and transient events in radio astronomy images \citep{Hancock_Robbie_2018}.
\robbie{} processes data in the following way.
First, the individual epoch images are warped to remove any astrometric distortions introduced by the ionosphere, which are not captured in the calibration stage.
This warping is done using \fitswarp{}\footnote{www.github.com/nhurleywalker/fits\_warp} \citep{2018A&C....25...94H}.
Next, the epochs are stacked to create an image cube, which is then flattened into a mean image.
This mean image is then used to create a reference catalogue of persistent sources.
For each of the persistent sources, \robbie{} creates a light curve using the so-called {\em priorized} fitting capability provided by \aegean{}\footnote{www.github.com/PaulHancock/Aegean} \citep{hancock_compact_2012,hancock_source_2018}.
Each of the light curves are then characterized using standard metrics including the de-biased modulation index $m_d$, $\chi^2$, and the false detection probability of seeing a given light curve variation given the uncertainties and a model of no variability.
Each of the individual epoch images are then masked using the persistent source catalogue, and blind source finding is run on each to generate a list of candidate transient events.
This list is then filtered for obvious false positives, and presented for further analysis.

We ran \robbie{} separately on each field/frequency combination making a total of six separate instances.
Sources were identified as variable if their light curves had $m_d>0.05$ and false detection probability of $\mathrm{p\_val} < 0.001$.
Four sources were identified as being variable in Field 5, with the variability being present at both frequencies.
Closer inspection of these light curves and images revealed that the cause of the variability was Jupiter moving through the field and passing close enough to persistent sources to become confused.
This detection of Jupiter highlights the effectiveness of \robbie{} at detecting variability, but means that no astronomical (extra-solar system) variability was observed.

We inspected the list of candidate transient events and found that all the candidates were in one of four categories: sources that were below the $5\sigma$ detection limit which were occasionally detected due to differing noise contributions, side-lobes of real sources that were not automatically excluded by \robbie{}, the Moon (present in a single epoch), or Jupiter when it was not being confused with persistent sources.
We therefore find no astronomical transients in any of the six datasets.

\section{Results}\label{sec:results}
As noted by \citet{bell_automated_2011} and others, the density of transients and variables can be calculated from  Poissonian statistics:
\begin{equation}
P(n) = \exp\left(-\rho \Omega\right)
\label{eq:rho}
\end{equation}
\noindent where $\rho$ is the surface density, $\Omega$ is the sky area surveyed, and n is the number of variable or transient events.
In the event of no detected events we can calculate the a $95\%$ confidence upper limit on the surface density by setting $P(n)=0.05$ and solving for $\rho$.
For multiple epochs and multiple pointing directions we can replace $\Omega$ with the equivalent sky area:
\begin{equation}
\Omega_{eq} = \sum \Omega_i N_i
\label{eq:omega}
\end{equation}
\noindent where $\Omega_i$ and $N_i$ are the sky coverage, and number of epochs for each pointing direction $i$.
The number of epochs, sky area, and upper limit on the density of variable sources for each field are tabulated in Table \ref{tab:var_results}.
The variable source population probed by this work are those brighter than the detection limit in the mean image produced by \robbie{}. 
Since the sensitivity changes over the field of view we report the mean sensitivity for the variable sources in Table\,\ref{tab:var_results}.
The upper limit on the transient sources population, however, needs to consider the maximum image noise among all epochs.
Thus we report a separate sensitivity for transient events in Table\,\ref{tab:var_results}.

We combined data from all three fields to produce a combined detection limit.
To do this, the effective area is equated as per Eq.\,\ref{eq:omega}, and we calculate the relevant sensitivity using:

\begin{equation}
    \mathrm{rms}_{Comb} = \frac{\sum \mathrm{rms}_i N_i}{\sum N_i}
    \label{eq:rms}
\end{equation}

\begin{table}[h]
\centering
\begin{tabular}{cccccccc|cc}
\hline
\multicolumn{8}{c|}{Measured} & \multicolumn{2}{c}{Expected}\\
Field & $N_{epoch}$ & Freq. & $N_{src}$ & $\Omega_i$    & rms$^{V}$  & rms$^{T}$ & $\rho$     & $\bar{m}$ & $\bar{\tau}_0$\\
      &             & (MHz) &           & deg$^2$ & (mJy)      & (mJy)     & deg$^{-2}$ & \%        & year\\
\hline
3    & 9 & 154 & 2625 & 1028 & 88  & 230 &  $<3.2\times 10^{-4}$ & 16 & 2.9\\
3    & 9 & 185 & 2594 & 1028 & 65  & 380 &  $<3.2\times 10^{-4}$ & 17 & 1.9\\
\hline
4    & 2 & 154 & 1609 & 1028 & 225 & 1700 &  $<1.5\times 10^{-3}$ & 11 & 9.3\\
4    & 2 & 185 & 2168 & 1028 & 79  & 715 & $<1.5\times 10^{-3}$ & 12 & 6.2\\
\hline
5    & 6 & 154 & 1814 & 718  & 78  & 192 &  $<6.9\times 10^{-4}$ & 12 & 5.1\\
5    & 5 & 185 & 1628 & 718  & 65  & 289 &  $<8.3\times 10^{-4}$ & 14 & 3.4\\
\hline
\end{tabular}
\caption{Summary of data for the three fields of interest.
$N_{src}$ is the number of persistent sources detected.
$\Omega_i$ is the sky area used in Eq.\,\ref{eq:omega}.
The sensitivity limit for variable sources (rms$^{V}$) and transient sources (rms$^{T}$) are separately reported as described in the text.
$\rho$ is a limit on the surface density of events detected in the survey as per Eq.\,\ref{eq:rho}, and is the same for variable and transient events since none were detected.
The final two columns show the expected modulation index $\bar m$ and characteristic timescale $\bar{\tau}_0$ for refractive interstellar scintillation and are based on the {\sc SM2017} model.}
\label{tab:var_results}
\end{table}

In order to assess the implications of the data in Table\,\ref{tab:var_results}, we must calculate the expected variability rate.
The overwhelming majority of sources that are detected by the MWA are AGN.
The characteristic timescale for AGN intrinsic variability is on the order of years to decades.
Since the observations of the K2 fields were conducted on time scales of days to weeks, we do not expect to see any intrinsic variability.
Extrinsic variability, in the form of strong refractive interstellar scintillation (RISS), can be calculated using a model of scintillation based on Galactic $H_\alpha$ intensity.
Specifically, we use $H\alpha$ intensity from \citet{Haffner_faint_1998} as a proxy for scattering measure following their Eq 1. and Eq 16. of \citet{Cordes_NE2001_2002}.
This scattering measure is then used to calculate the diffractive scale using Eq 7a of \citet{Macquart_temporal_2013}.
These calculations are carried out using the {\sc SM2017}\footnote{www.github.com/PaulHancock/SM2017/} code, which assumes a scattering screen distance of 1\,kpc.
The modulation index and timescale of variability is calculated for each of the persistent sources in each of the fields at each frequency, and we report the mean of these in Table\,\ref{tab:var_results}.
The expected modulation indexes are $10-15\%$, which should be easily detectable given our selection criteria of $m_d > 5\%$.
However the timescale of variability for RISS is $1-10$ years, whereas our observations only cover timescales of days to weeks.
The amount or variability seen on these shorter timescales would therefore be only $\sim 1/300^{\mathrm{th}}$ of the raw modulation index, below our ability to measure.
The low expectation for variability, and the lack of observed variability are consistent corresponding to a low false detection rate for any transient sources found.

\section{Discussion and Conclusions}
\begin{figure}
    \centering
    \includegraphics[width=0.9\linewidth]{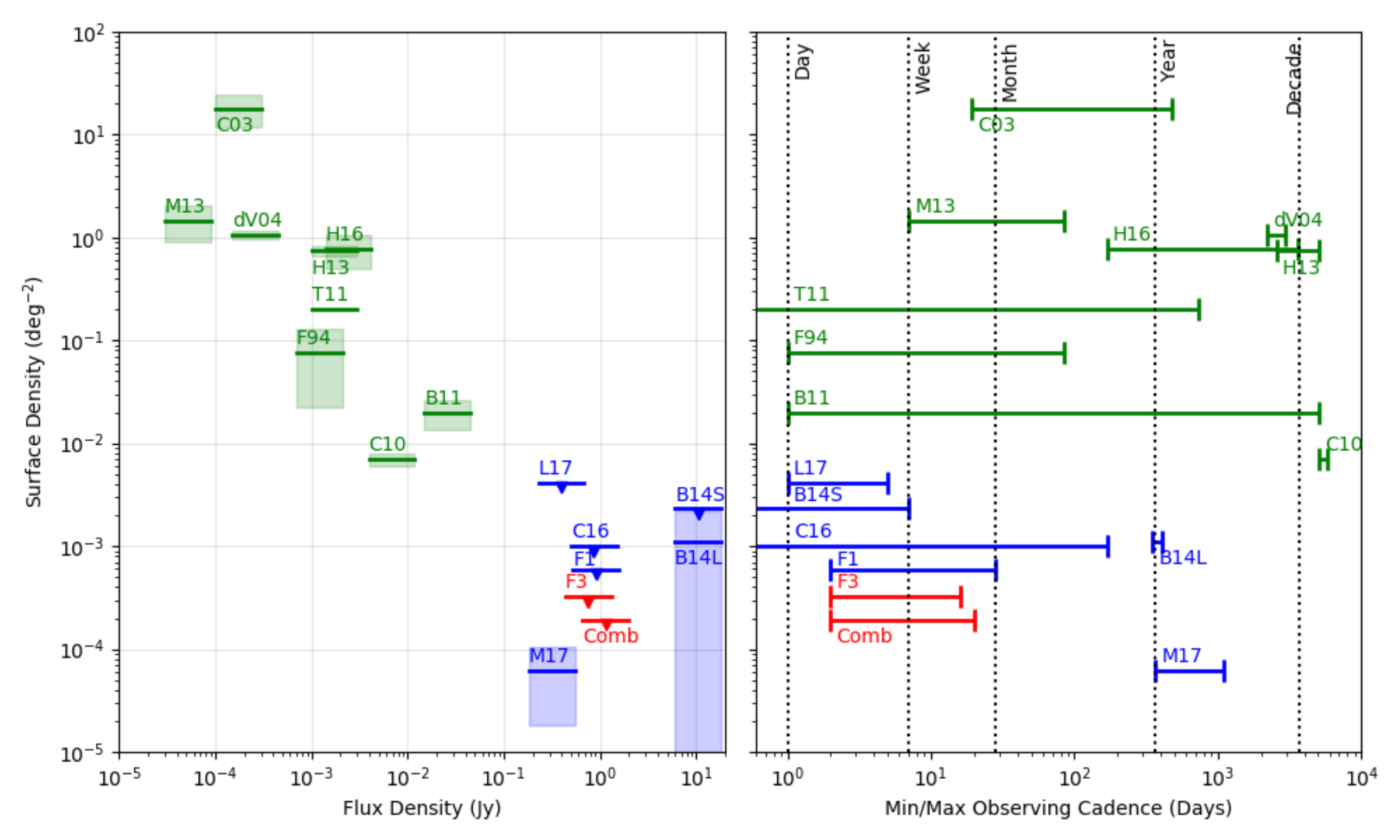}
    \caption{
    {\bf Left}: Areal source density of variable sources as a function of survey sensitivity for surveys at $\sim 1$\,GHz (green), and $\sim 150$\,MHz (blue), as well as the current work at $154$\,MHz (red).
    Horizontal lines represent $1-3$ times the detection limit, vertical shaded areas are uncertainties based on the number of detections, and triangles indicate upper limits.
    F3 is the upper limit from field 3 of this work at $154$\,MHz (the most sensitive individual limit), and Comb is the combination of fields $3-5$ at the same frequency.
    {\bf Right}: The min/max observing cadence for the same surveys, with colors as per the left panel.
    In this plot the surface density measurements and limits should vary as a function of time scale but that has not been taken into account.
    {\bf References}: $1$\,GHz data: C03 \citet{carilli_variability_2003}, M13 \citet{mooley_sensitive_2013}, dV04 \citet{de_vries_optical_2004}, H13 \citet{hodge_millijansky_2013}, H16 \citet{hancock_radio_2016}, T11 \citet{thyagarajan_variable_2011}, F94 \citet{frail_search_1994}, B11 \citet{bannister_22-yr_2011}, C10 \citet{croft_allen_2010}. $150$\,MHz data: M17 \citet{2017MNRAS.466.1944M}, L17 stokes-I limits from \citet{lynch_search_2017} on a daily cadence, C16 \citep{carbone_new_2016}, F1 Kepler field 1 from \citet{PaperI}, B14S is the short duration limit and B14L is the long duration detection of variability from \citet{2014MNRAS.438..352B}.
    }
    \label{fig:lognlogs}
\end{figure}

In Figure\,\ref{fig:lognlogs} we show the areal source density as a function of sensitivity for a selection of surveys at $\sim 1$\,GHz and $154$\,MHz.
Also indicated is the minimum and maximum timescale that each survey is sensitive to.

In comparison to other radio surveys for variable and transient events (Figure\,\ref{fig:lognlogs}), we find a significantly smaller surface density for each.
Our upper limit on the density of variables is an order of magnitude lower than that found at comparable sensitivities by \citet{bannister_22-yr_2011} and \citet{croft_allen_2010} at higher frequencies ($\sim 1$\,GHz).
We attribute this discrepancy to two effects: the difference in observing frequency; and the cadence of observations.
The \citet{bannister_22-yr_2011} and \citet{croft_allen_2010} surveys were most sensitive to variability on the timescales of years to decades, whereas this work samples days to weeks.
As predicted by the SM2017 model, we would expect RISS to have characteristic timescales of years to decades as per Table\,\ref{tab:var_results}.

Also shown in Figure\,\ref{fig:lognlogs} are surveys at $154$\,MHz.
\citet{2014MNRAS.438..352B} used the MWA commissioning array to produce a survey with a $5\sigma$ sensitivity of 5.5\,Jy and a cadence of either less than 10 days or a year.
They detect a single variable source giving a surface density of variables of $1\times 10^{-3}\,\mathrm{deg}^{-2}$ on year long time-scales, but find no variables on the shorter time-scales.

\begin{figure}
    \centering
    \includegraphics[width=0.45\linewidth]{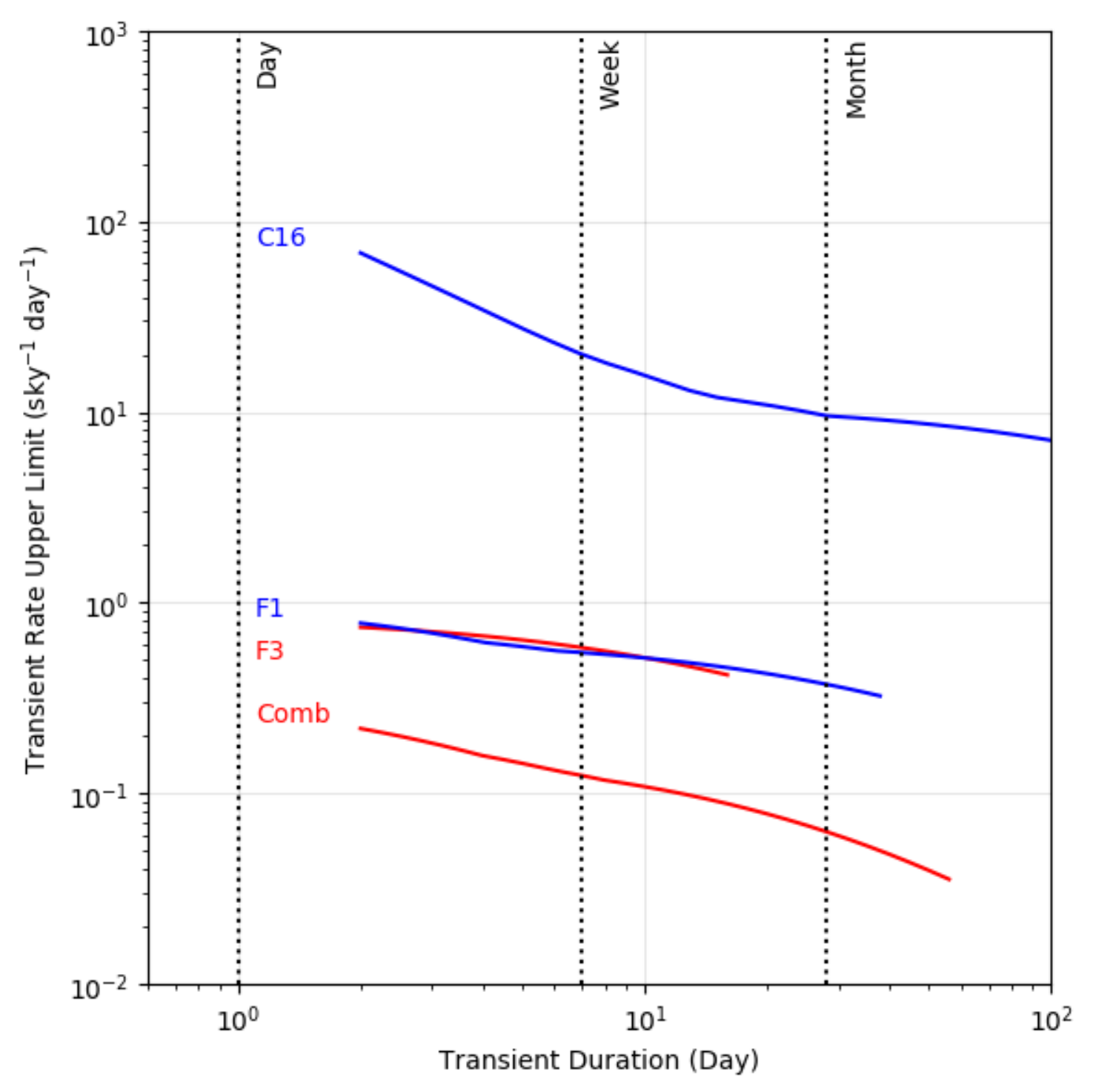}
    \caption{The transient rate upper limit for four of the surveys shown in Figure\,\ref{fig:lognlogs}, for transients longer than 2 days. The upper limits take into account the distribution of observing epochs and the likelihood that a transient can occur between observations, as per the formalism of \citet{carbone_new_2016}.
    The curves for F1 and F3 are nearly identical over $2-16$\,days.}
    \label{fig:carbone}
\end{figure}

\citet{2017MNRAS.466.1944M} compare the TIFR GMRT Sky Survey Alternative Data Release 1 \citep[TGSS ADR1,][]{intema_gmrt_2017} and the GaLactic and Extragalactic All-sky Murchison Widefield Array \citep[GLEAM,][]{hurley-walker_galactic_2017} survey catalogues and find a single variable source between two epochs separated by $1-3$ years.
However \citet{lynch_search_2017} found no variable sources in their total intensity images created from nightly observations, and \citet{PaperI} found no variability in Kepler Field 1 on time-scales of days to a month.  Shown in Figure\,\ref{fig:lognlogs}, the combination of our Fields 3, 4, and 5 results provide the lowest upper limit ($2\times10^{-4}$\,deg$^{-2}$) at these frequencies, sensitivities ($660$\,mJy), and cadences (day-month) from blind surveys.

\citet{carbone_new_2016} used 37 Dutch stations of LOFAR with a sub-set of the high band array, to survey four fields over a period of 6 months. They found no variable or transient sources, and their reported upper limit is shown as C16 in Figure\,\ref{fig:lognlogs}.
Additionally, \citet{carbone_new_2016} present a new method to represent the survey sensitivity for transient events as a function of the transient duration. In Figure\,\ref{fig:carbone} we replicate their results and add those for the Kepler fields F1, F3, and Comb, as described in Figure\,\ref{fig:lognlogs}.
The combined data set indicates a much lower limit on the rate of transients on timescales of 2 days (0.2\,sky$^{-1}$day$^{-1}$) to 1 month (0.06\,sky$^{-1}$day$^{-1}$).
The upper limits for the transient rate of F1 and F3 in Figure\,\ref{fig:carbone}, are nearly identical over the $2-16$\,day range. F1 has a smaller survey area than F3, but a greater number of epochs, and the these two factors cancel out.
The C16 survey has approximately 10 times the number of epochs as the F1/F3 surveys, but a field of view that is nearly $1/70th$ the size, resulting in a much less stringent upper limit.
For transients of duration $2-16$\,days, the combined data of all three fields from this paper (Comb) has an effective field of view that is 3 times that of the individual surveys, and a similar distribution of epochs, resulting in a more stringent limit on the transient rate.
In all, Figure\,\ref{fig:carbone} demonstrates the advantage of a wide field of view instrument such as the MWA with it's ability to achieve high sensitivity to transient events in only a small number of observations.

Additional low frequency surveys are underway with the GMRT\footnote{\href{http://www.tauceti.caltech.edu/stripe82/g1sts/}{www.tauceti.caltech.edu/stripe82/g1sts/}} and LOFAR \citep{fender_lofar_2006}, however they are yet to be published.

The discrepancy between the detection and non-detections listed here can be explained by the different phase space that they are exploring: long versus short timescale variability.
At $1$\,GHz variability is seen on timescales of days to decades, to varying degrees, where as at $154$\,MHz the variability is only seen on year long time-scales.
This difference is consistent with the $H\alpha$ based modeling presented in the previous section, and means that at these frequencies intrinsic incoherent variability from AGN are on similar timescales to the extrinsic RISS induced variability, making the two difficult to disentangle.
However, the corollary is that short time-scale variability must be dominated by other effects such as ionospheric and instrumental effects, but also intrinsic coherent emission.
This makes the quiet incoherent low frequency sky an ideal place to look for coherent variability including: stellar flares \citep{lynch_154_2017}; cyclotron maser emission from extra-solar planets \citep{farrell_radio_2004,sirothia_search_2014}; prompt emission from gravitational wave events \citep{chu_capturing_2016}; fast radio bursts \citep{2016MNRAS.458.3506R, 2015AJ....150..199T}; and new classes of as-yet unidentified Galactic transients \citep{2005Natur.434...50H,2012AJ....143...96J}.
As such we anticipate that low-frequency surveys such as the GMRT 150 MHz Stripe 82 Transient Survey\footnote{\href{http://www.tauceti.caltech.edu/stripe82/g1sts/}{www.tauceti.caltech.edu/stripe82/g1sts/}}, the LOFAR Transients Key Science Project \citep{fender_lofar_2006}, and MWA Transients Survey full data release \citep{2019MNRAS.482.2484B}, will be moving into a new area of discovery space that is more closely linked to the physics of the emission process rather than the physics of the interstellar medium.  And in the more distant future, these surveys will be pathfinders for low frequency transient/variable surveys with the SKA.

\section*{Acknowledgments}

\subsection*{Software}
We acknowledge the work and support of the developers of the following following software: Astropy \citet{the_astropy_collaboration_astropy_2013,astropy_collaboration_astropy_2018}, Numpy \citep{van_der_walt_numpy_2011}, Scipy \citep{Jones_scipy_2001}, Pandas \citep{mckinney_pandas_2010}, AegeanTools \citep{hancock_source_2018}, Topcat \citep{Taylor_topcat_2005}, \fitswarp{} \citep{2018A&C....25...94H},
 DS9\footnote{\href{http://ds9.si.edu/site/Home.html}{ds9.si.edu}}, \robbie{} \citep{Hancock_Robbie_2018}, and SM2017\footnote{\href{https://www.github.com/PaulHancock/SM2017/}{www.github.com/PaulHancock/SM2017/}}.


This research made use of Astropy, a community-developed core Python package for Astronomy \citep{the_astropy_collaboration_astropy_2013,astropy_collaboration_astropy_2018}.

\subsection*{Facilities}

This scientific work makes use of the Murchison Radio-astronomy Observatory, operated by CSIRO. We acknowledge the Wajarri Yamatji people as the traditional owners of the Observatory site. Support for the operation of the MWA is provided by the Australian Government (NCRIS), under a contract to Curtin University administered by Astronomy Australia Limited. We acknowledge the Pawsey Supercomputing Centre which is supported by the Western Australian and Australian Governments.
This research has made use of NASA's Astrophysics Data System.

\facility{Murchison Widefield Array}

\bibliography{biblio} 

\end{document}